\documentclass[12pt,twoside]{article}
\usepackage{xrb2007}
\usepackage{psfig}
\newcommand\mdotlong {\hbox{$\langle \dot{M} \rangle$}}

\newcommand\mdot {\hbox{${\dot M}$}}
\newcommand\mpy {\hbox{$M_\odot~{\rm yr}^{-1}$}}

\begin{document}

% select your session by uncommenting the appropriate line
%\session{Jets}
%\session{Jet and Black Hole Binaries}
%\session{Faint Galactic XRB Populations}
%\session{Faint XRBs and Galactic LMXBs}
%\session{Obscured XRBs and INTEGRAL Sources}
%\session{ULXs}
%\session{Extragalactic Populations}
%\session{Future Missions and Surveys}
\session{Population Synthesis}

\shortauthor{R. Wijnands}
\shorttitle{Sub-luminous accreting neutron stars}

\title{Enigmatic sub-luminous accreting neutron stars
in our Galaxy}

\author{Rudy Wijnands}
\affil{Astronomical Institute 'Anton Pannekoek', University of Amsterdam, 
Kruislaan 403, 1098~SJ, Amsterdam, The Netherlands}

\begin{abstract}
During the last few years a class of enigmatic sub-luminous accreting
neutron stars has been found in our Galaxy. They have peak X-ray
luminosities (2--10 keV) of a few times $10^{34}$ erg s$^{-1}$ to a
few times $10^{35}$ erg s$^{-1}$, and both persistent and transient
sources have been found. I present a short overview of our
knowledge of these systems and what we can learn from them.
\end{abstract}

\section{Sub-luminous accreting neutron stars in low-mass X-ray binaries}

Instrumental limitations (e.g., sensitivity, spatial resolution) of
past X-ray instruments restricted the study of accreting neutron stars
in low-mass X-ray binaries to the relatively bright systems, with
2--10 keV X-ray luminosities of $10^{36-38}$ erg s$^{-1}$. The
existence of fainter sources was well known, but until recently they
could not be studied in detail. Luckily, this is not true for the
current X-ray satellites and a sizable number ($\sim$15) of
sub-luminous (X-ray luminosities of $10^{34-35}$ erg s$^{-1}$)
accreting neutron stars have been found in the last few years (called
sub-luminous NS LMXBs; I will only consider the systems in which the
companion is suspected to have a low mass [$<$1 $M_\odot$] but there
exist also a population of systems which have a suspected high-mass
companion). Their neutron star nature has been confirmed by the
detection of type-I X-ray bursts.  Two of them are persistent sources,
while the others are transients with a large variety in outburst
durations and recurrence times.\

\vspace{0.2cm}

\noindent
{\bf The persistent sub-luminous NS LMXBs:} Currently, two such
sources are known: 1RXS J171824.2--402934
\citep{1999A&A...349..389V,2000A&A...358L..71K} 
was shown to be persistent with a 0.5--10 keV X-ray luminosity of only
$\sim 5\times 10^{34}$ erg s$^{-1}$ by \cite{2005A&A...440..287I}.
After AX J1754.2--2754 \citep{2002ApJS..138...19S,
2007ATel.1094....1C} exhibited a type-I X-ray burst,
\cite{2007ATel.1143....1D} detected it with Swift at a
2--10 keV luminosity of a few times $10^{34}$ erg s$^{-1}$
establishing it too as a persistent sub-luminous NS LMXB.\

\vspace{0.2cm}

\noindent
{\bf The quasi-persistent sources:} Recently, two quasi-persistent
(outburst durations of years) sub-luminous NS LMXBs have been
identified. The best example is XMMU J174716.1--281048
\citep{2003ATel..147....1S}. \cite{2006ATel..970....1B}, using
INTEGRAL, reported on a new type-I X-ray burster near the Galactic
center which was very soon identified with XMMU J174716.1--281048
\citep{2006ATel..972....1W}. After analyzing the burst properties and
several XMM-Newton observations, \cite{2007A&A...468L..17D} identified
the source as a quasi-persistent sub-luminous NS LMXB with a
luminosity of $\sim 5 \times 10^{34}$ erg s$^{-1}$; a classification
which was later confirmed by a series of Swift and Chandra observation
of the source in 2007
\citep{2007ATel.1078....1D,2007ATel.1136....1D,2007ATel.1174....1S}.

\cite{1996PASJ...48..417M} first discovered the second system, AX
J1745.6--2901, as a bursting transient near Sgr A*, with a luminosity
of a few times $10^{35}$ erg s$^{-1}$ and with deep eclipses every
$\sim$8.4 hours (the orbital period). They suggested that AX
J1745.6--2901 could be the old, bright transient A 1742--289 (albeit
this time in a faint outburst), but
\cite{1996PASJ...48L.117K} could not find eclipses in the
Ariel V data of A 1742--289 indicating that AX J1745.6--2901 is a new
transient sub-luminous NS LMXB. Quasi-daily observations of the
Galactic center with Swift in February 2006 yielded a new transient,
Swift J174535.5--290135.6 \citep{2006ATel..753....1K}, which
XMM-Newton conclusively identified with AX J1745.6--2901
\citep[showing eclipses with the right
period;][]{2007ATel.1058....1P}.  The source turned of after staying
on for several months \citep{deegwijnands}, but it was detected in
outburst again in February 2007
\citep{2007ATel.1005....1K,2007ATel.1006....1W}.  The Swift
quasi-daily monitoring campaign continued also in 2007 showing that
the source remained active throughout 2007
\citep{deegwijnands}.\

\vspace{0.2cm}
\noindent
{\bf The transient sources:} In addition, about a dozen,
short-duration, transient sub-luminous NS LMXBs are known \citep[see,
e.g.,][]{2006A&A...449.1117W}. In general, they have peak outburst
luminosities (2--10 keV) of $10^{34-35}$ erg s$^{-1}$ and their
outbursts last from a few days to at most a few weeks. The combination
of their very-faint peak luminosity and their transient nature makes
them difficult to discover and study, but dedicated programs
\citep[e.g.,][]{2006A&A...449.1117W} are increasing the number of
these systems known and our understanding of them.

\section{What can we learn from them}

The total number of sub-luminous NS LMXBs in our Galaxy is unknown.
That they are found in ever growing numbers despite the difficulties
in observing them suggests that they might be quite abundant.  Binary
evolution and population synthesis models might need modifications to
explain their properties \citep[e.g., unusual binary configuration,
see e.g.,][]{2005A&A...440..287I} and their number density in our
Galaxy. Also, these systems may help us understand the effects of
accretion on super-dense neutron-star matter as I explain below.\

\begin{figure}
\hbox{\psfig{figure=wijnands_r_fig1.ps,height=6.5cm}~~~~~~~~
~~~~~~~~~\psfig{figure=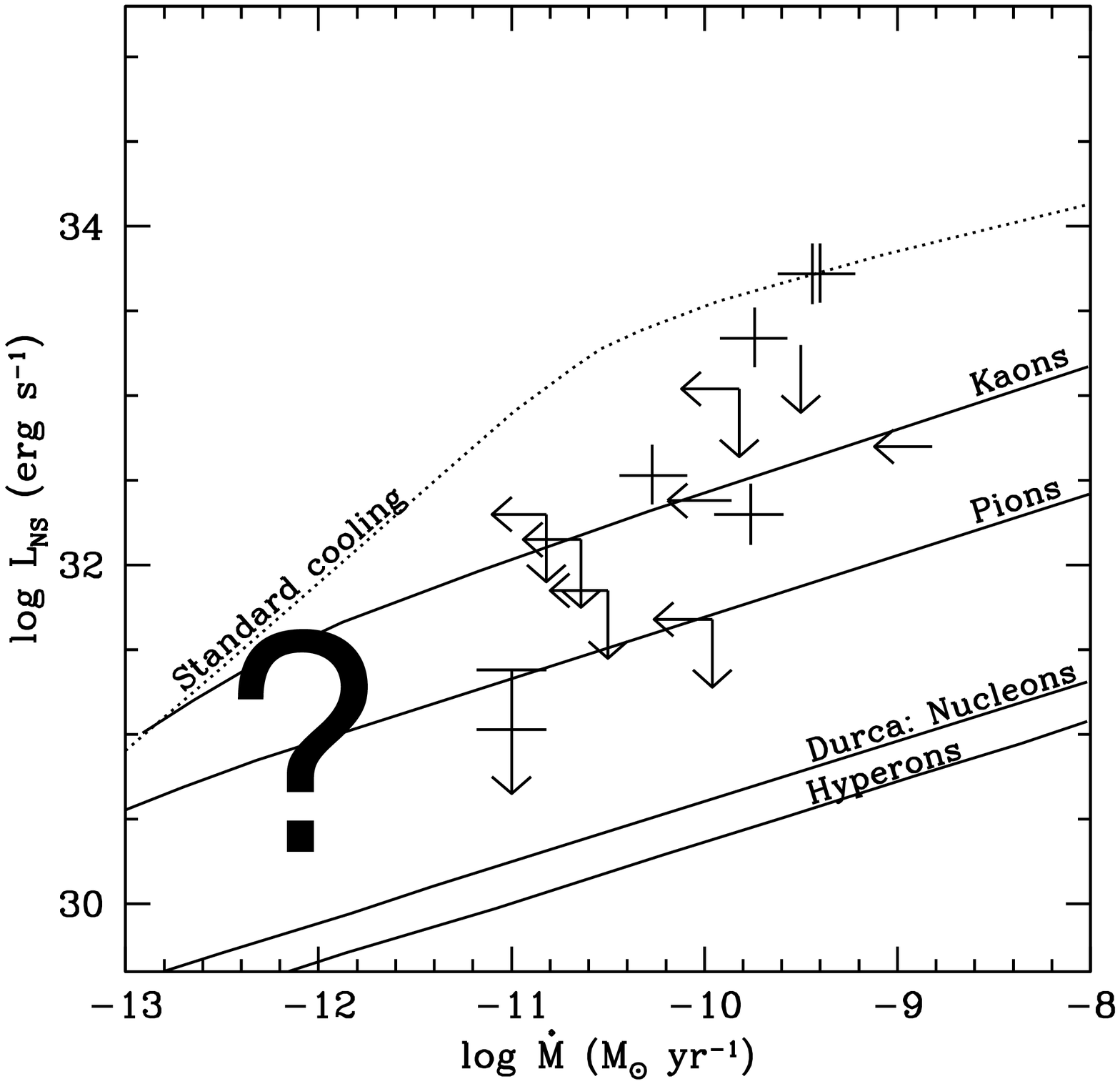,height=7cm}}
\caption{Left: the number of AMXPs versus 
\mdotlong~\citep[excluding Aql X-1;
after][]{2006AIPC..840...50G}.  Right: luminosities of quiescent
neutron-star transients versus
\mdotlong~\citep[after][]{2007ApJ...660.1424H}. Question marks
indicate where the sub-luminous NS transients are
expected. \label{fig}}
\end{figure}

\vspace{0.2cm}
\noindent
{\bf Accreting millisecond X-ray pulsars:} It is thought that NS LMXBs
are the progenitors of the millisecond radio pulsars, thus, accreting
systems must also harbor a fast spinning neutron star. The 9 known
accreting millisecond X-ray pulsars \citep[AMXPs; excluding Aql X-1
whose pulsations are not conclusively accretion
driven;][]{2007arXiv0708.1110C} confirm this hypothesis but the
question remains: why are the other systems not AMXPs?  It has been
suggested that in the non-pulsating systems the neutron-star magnetic
field could be temporarily buried by the accreted matter. When the
accretion phase ends, the field reemerges from the crust and the
systems become millisecond radio pulsars. If the time-averaged mass
accretion rate \mdotlong~of the accretors is relatively high, the
accreted matter can bury the field throughout the life of the X-ray
binary making it impossible to see pulsations. However, if
\mdotlong~is low enough for the field to diffuse out of the accreted
matter faster than it is buried then these systems will exhibit
pulsations \citep[e.g.,][]{2001ApJ...557..958C}.  The exact processes
behind field-burying depend on the properties of the neutron-star
crust, which in turn depend on the equation of state of neutron-star
matter and how the accreted matter burns on the surface of the star
(determining the composition of the crust). Seven of the known
AMXPs have indeed among the lowest~\mdotlong~of the relatively bright
transients (Fig.~\ref{fig}, left), pointing to the importance
of~\mdotlong~in this process. Two systems have higher
\mdotlong~(resembling that of the non-pulsating systems) but their
pulsations were intermittent and possibly triggered by bright
X-ray flashes \citep{2007ApJ...654L..73G,2007arXiv0708.1316A} which
might have disturbed the screening currents in the crust, temporarily
letting the field out before it was buried again by the accreted
matter. This also suggests that field-burying is relevant in
understanding the presence or absence of pulsations in neutron-star
accretors.

Unfortunately, firm conclusions are not possible with only 9
AMXPs. Luckily, the sub-luminous NS LMXB can help because they have
similar (the persistent sources) or even lower \mdotlong~(the
transients).  If truly all neutron stars have an intrinsic magnetic
field and this field is indeed buried in the non-pulsating systems,
then the sub-luminous NS LMXBs should also be visible as AMXPs because
their field cannot be buried.  If most sub-luminous NS LMXBs are found
to be AMXPs the field-burying scenario is strengthened, plus, a larger
number of AMXPs enable more detailed comparative studies between
pulsating and non-pulsating systems to elucidate the field-burying
physics and provide new insights into the behavior of matter and
magnetic fields under extreme conditions. But if only sporadically an
AMXP is found, then a mechanism other than field-burying is
responsible for preventing observable pulsations for most NS LMXBs,
and we are back to the drawing board! \

\vspace{0.2cm}
\noindent
{\bf Thermonuclear flashes:} The properties of type-I X-ray bursts
(rise and decay times, energetics, recurrence times) depend strongly
on the characteristics of the neutron star (i.e., in the crust) and
the chemical composition of burning material. The latter can change
drastically as the \mdot~onto the neutron star varies, defining
several burning regimes which lead to different types of flashes
\citep[see][]{1981ApJ...247..267F,1998mfns.conf..419B}.  Although the exact
\mdot~range for each regime depends on the specifics of the
X-ray flash model used and on the assumed properties of the neutron
star (e.g., crust microphysics, compactness, core temperature), most
models roughly agree on the values of these limits. However,
observationally, some sources fit the predictions while others deviate
significantly \citep[see review
of][]{2006csxs.book..113S}. Nevertheless, a study of $\sim1000$
flashes from many sources using the RXTE archive
clearly established that the great majority of sources
behave consistently with the theoretical~\mdot~limits
\citep{2006astro.ph..8259G}. Thus, although the details of the physics
behind flashes still elude us, the global behavior is reasonably well
understood.

Unfortunately, this is really only true for the relatively fast
accretors (\mdot~$>$$10^{-10}$ \mpy).  Most of the flashes reported by
\citet{2006astro.ph..8259G} occurred when the neutron stars were
accreting at high rates, and while this study found that the flashes
observed in the lowest \mdot~regime behaved as predicted, most of the
flashes observed at the lowest~\mdot~come from one particular source
(EXO 0748--676) and firm conclusions cannot be drawn.  Similar results
were reported for a few other
sources~\citep{2002A&A...392..885C,2005A&A...440..287I}, but several
other systems \citep[not studied by][ because RXTE did not
observe them]{2006astro.ph..8259G} are not consistent with this
picture~\citep{2002A&A...392..885C}.  Recently,
\cite{2007ApJ...654.1022P} suggested an explanation: when the~\mdot~is
very low, the helium and CNO elements in the accreted matter
sediment out of the accreted fuel before it reaches the conditions at
which hydrogen ignites resulting in different flash behavior; when
\mdot~is high sedimentation is not important
\citep[see also][]{2007ApJ...661..468C}.  The limited observational
data precludes detailed testing of this scenario, and this model does
not explain the differences between the sources. Clearly, studying
type-I bursts from sub-luminous NS LMXBs will help us understand burst
physics and neutron-star reacts to the accretion of matter.\

\vspace{0.2cm}
\noindent
{\bf Cooling neutron stars:} The transient NS LMXBs are also
interesting when in their quiescent state, in which they are expected
to emit thermal surface radiation because the accreted matter heats
the star during the outburst \citep{1998ApJ...504L..95B}. The amount
of radiation depends on \mdotlong~and on the core cooling
processes. The latter depend strongly on the core composition; exotic
matter (e.g., kaon/pion condensates, hyperons, unbound quarks) would
cool the core much faster than if it was not present \citep[see review
by][]{2004ARA&A..42..169Y} and the curves in Fig.~\ref{fig},
right). By combining the observed emission with estimates of
\mdotlong, the core properties and thus the behavior of ultra-dense
matter can be probed.  Until now all studies involved transients with
\mdotlong~$\geq 10^{-11}$~\mpy, which showed that several of them have
relatively hot neutron stars which cool down without enhanced cooling,
but a large number have cores colder than expected if only standard
core cooling occurs
\citep[][Fig.~\ref{fig}, right]{1998ApJ...504L..95B}. Such a cold
core might result because enough matter was accreted onto the star to
increase its mass and core density to the point where enhanced core
cooling processes can occur
\citep{1998ApJ...504L..95B,2001ApJ...548L.175C,2004ARA&A..42..169Y}.

So far, the sub-luminous NS LMXBs with~\mdotlong~$\approx 10^{-12}$
\mpy~have not been studied in detail in quiescence, thus, it is
unclear what to expect from their quiescent thermal emission. If their
current levels of~\mdotlong~are representative of their whole
accretion life \citep[possible if they have primordial brown dwarf or
even planetary companions,][]{2006MNRAS.366L..31K} then they could not
have accreted enough matter to raise the core density above the
threshold for fast cooling processes (assuming all neutron stars are
born with roughly similar masses) and they should follow the 'standard
cooling' curve in Figure~\ref{fig} (right) and have thermal emission
easily detectable with Chandra or XMM-Newton. Confirming
this observationally would reaffirm the cooling models, falsifying it
could imply that fast cooling is also possible at moderate central
densities, or, more likely, that these systems may have accreted in
the past at levels higher than we observe now thereby accreting enough
matter to allow for enhanced core cooling processes. Detailed binary
evolution studies are needed to be conclusive.

Intriguingly, there is a class of sources which can significantly
contribute to core cooling studies in accreting neutron stars:
low-luminosity neutron-star X-ray sources in Galactic globular
clusters. More than a hundred have been identified as neutron-star
transients in their quiescent states with Chandra and XMM-Newton
\citep{2003ApJ...598..501H,2003ApJ...591L.131P,2006AdSpR..38.2930W}
based on their X-ray luminosities and thermal spectral shape.  This
identification must still be confirmed (i.e., by observing accretion
outbursts), but it is likely to be correct since no other known X-ray
sources have these properties and they are too abundant to be a
previously unknown object type.  Thus, we know their quiescent
luminosities accurately, but not their \mdotlong~because no accretion
outbursts have been observed from them. Using the limits on both peak
outburst luminosities and recurrence time provided by monitoring X-ray
instruments and the many sensitive pointed X-ray observations of the
Galactic globular clusters, I estimate an averaged \mdotlong~for these
systems of typically $10^{-12}$~\mpy~or less, precisely in the range
of the sub-luminous NS LMXBs. Plotting this \mdotlong~in
Figure~\ref{fig} (right) with the measured quiescent luminosities,
these systems can only be explained if standard core cooling alone
occurs in their cores. Sensitive monitoring observations of a large
number of globular cluster systems should be performed to catch the
expected very-faint X-ray outbursts which would conclusively establish
them as NS LMXBs.

\end{document}